# Vector beam generation from standing hollow GaN nanowire lasers on sapphire substrate

Masato Takiguchi[1,2*], Sylvain Sergent[1,2], Benjamin Damilano[3], Stéphane Vézian[3], Sébastien Chenot[3], Nicole Yazigi[3], Peter Heidt[2], Tai Tsuchizawa[1,4], Taiki Yoda[2,5], Hisashi Sumikura[1,2], Akihiko Shinya[1,2], and Masaya Notomi[1,2,5]

[1]Nanophotonics Center, NTT Corp., 3-1, Morinosato Wakamiya, Atsugi, Kanagawa 243-0198, Japan

[2]NTT Basic Research Laboratories, NTT Corp., 3-1, Morinosato Wakamiya, Atsugi, Kanagawa 243-0198, Japan

[3]Université Côte d'Azur, CNRS, CRHEA, Rue B. Grégory, Valbonne 06560, France

[4]NTT Device Technology Laboratories, NTT Corp., 3-1, Morinosato Wakamiya, Atsugi, Kanagawa 243-0198, Japan

[5]Department of Physics, Tokyo Institute of Technology, 2-12-1 Ookayama, Meguro-ku, Tokyo 152-8550, Japan

## Abstract

We fabricated GaN based hollow nanowires standing upright on a sapphire substrate by the sublimation method and found that they exhibit laser oscillation at room temperature. These very long, hollow, nano-sized structures cannot be fabricated by other means. Furthermore, we determined the condition under which the fundamental mode is azimuthally polarized by investigating the dispersion of the hollow structure. Examination of the measured emission properties indicates that the hollow nanowire operates as a topological, vector-beam, light source.

## Introduction

Topological light[1] is classified into orbital angular momentum (OAM) light and axially polarized (vector-beam) light. Both types have singular points of phase and polarization on the optical axis, so they are considered to be potentially useful in singular optics[2]. Their potential applications include optical communications[3,4], quantum information processing[5], laser cutting[6], and trapping of particles[7] and atoms[8]. However, topological light has been mainly studied in free space optics, and there are no reports on topological emitters based on nanomaterials or their integration into optical circuits. Thus, realization of such miniaturized light sources is very significant. In particular, if topological light sources can be miniaturized and integrated densely on-chip, the new degrees of freedom they offer might be exploited for optical computing, while nano-sized particle trapping devices could be useful for sensing applications[9,10,11].

Our study focused on generating axially polarized beams (vector-beams). Here, various generation methods have been proposed. A simple method is mode manipulation by stressing optical fiber from the outside[12,13], but it suffers from mode controlling difficulties and optical loss. Even when a ring resonator[14,15], a plasmonic metasurface[16,17] or a hologram[18,19] is used, it is a passive element and requires an external light source. Of the methods that use active devices, a simple one is to create point defects in a Fabry-Perot cavity[20,21]. This method is very inefficient because it introduces loss in the cavity, and the device itself may be bulky, which increases its operation energy. Recently, signal processing of vector beams using a ring resonator laser has been demonstrated[22]. This method generates radially polarized light and is capable of fast signal modulation. Its drawback is a large footprint.

We investigated the use of GaN hollow nanowire lasers standing on sapphire substrate as tiny generators of vector beams with azimuthal polarization (Figure 1 and Figure 2). Nanowires[23,24,25] have promising optical and electrical characteristics. Various nanowire light sources[26,27] have been developed so far. However, there is no previous report on the fabrication of hollow nanowires capable of efficient vector-beam generation. In addition, various materials have been used to fabricate small hollow structures, but none have been usable as lasers. For example, carbon nanotubes[28,29,30] are hollow but their quantum efficiency is too low for them to be used as emitters. Small lasers with hollow structures have been demonstrated in the form of micrometer-sized rolled-up semiconductor emitters and lasers[31,32]. However, they are much larger than semiconductor nanowires and still too big to be integrated into optical circuits as nanodevices. Recently, vector beams have been generated from nanowires with a donut shaped reflector at the bottom under cryogenic temperature[33,34]. In particular, it has been shown that nanowires support multimode propagation, and the vector beam is generated from a higher order mode. Generally, however, the fundamental mode has a higher effective refractive index, and is advantageous for laser oscillation because more light is confined in the nanowire. Thus, conventional nanowires that rely on a higher order donut-shaped mode to generate the vector beam do so inefficiently. Our hollow nanowires have a fundamental donut-shaped electric field distribution ($TE_{01}$ mode), and allow us to realize vector beam lasers on substrates efficiently, even at room temperature. Although hollow nanowires ($Al_2O_3$ and AlInP) have been fabricated using wet etching methods[35,36], the former is a passive material, and the latter has poor optical gain (the optical properties were not reported). The hollow nanowires produced by our sublimation method have considerably better luminescence properties so they are advantageous for vector beam lasing at room temperature.

In regard to applications, development of waveguides for donut beams (optical fibers that also have an annular mode) has been very active[37,38]. If our light source can be coupled with such waveguides, the result could be used to make a signal transmission system. Moreover, our method can potentially implement lasers on Si substrate instead of sapphire and thus would be useful for making optoelectronic devices. Another potential application is optical trapping and optical manipulation. Here, nano-sized donut-shaped near field pattern (NFP) lasers would be able to create a circular electric field in a tiny space. Thus, our hollow nanowires capable of emitting topological light can contribute to new communication-related fundamental technologies and optical trap research fields.

## Fabrication

Our samples were fabricated by using the selective area sublimation method[39,40]. First, 4-µm-thick GaN on c-plane sapphire was grown by metal-organic vapor phase epitaxy. Next, 65-nm-thick $Si_xN_y$ was deposited by cathodic sputtering as a hard mask on top of the GaN layer. Then, 120-nm-thick negative-tone resist was spin-coated. A nanopattern was written into the resist by using electron-beam lithography and the patterns were transferred to the $Si_xN_y$ by using $SF_6$ (sulfur hexafluoride) reactive ion etching. After that, selective-area sublimation was carried out under ultrahigh vacuum in a molecular beam epitaxy reactor. We removed the SiN mask by HF wet etching. As shown by the SEM images in Figure 2, this fabrication method allowed us to form large aspect ratio vertical 4-µm-thick GaN nanowires on sapphire. Unlike the conventional nanowire fabrication method (self-catalyzed VLS approach, selective area growth, top-down plasma etching), our method is a metal-free process that allows us to fabricate arbitrarily shaped structures on sapphire with low sidewall damage.

It was not possible to determine whether the hole had penetrated the structure by just looking at its appearance with an SEM, so we directly observed the lengthwise cross section of one of the nanowires by cutting it with a focused ion beam (FIB). The SEM image of the cross section in supplemental 1 confirms that the hole penetrated all the way to the bottom of the wire.

Since sapphire has a lower refractive index than GaN, reflections can be expected to occur at the GaN/sapphire interface, forming a Fabry-Perot cavity. Many of the previous techniques could not make standing nanowires exhibiting laser oscillation because the refractive index of the growth substrate was the same as that of the nanowire itself or the refractive index of the substrate was higher. Therefore, it was necessary to narrow the space between the nanowires and the substrate[41]. However, our method allowed us to realize a nanowire laser standing on a substrate of lower refractive index material.

## Design

To determine the conditions under which a donut mode appears and the polarization, we investigated the dispersion relation of the waveguide mode in 2D-FEM simulated GaN nanowires with a hexagonal cross section. Figure 3 (a) shows the dispersion relation, where the x-axis is nanowire radius and the y-axis is effective refractive index. The dispersion relationship shown here is the same as that of a typical cylindrical waveguide, such as an optical fiber. The fundamental mode, called the $HE_{11}$ mode, has a Gaussian intensity distribution (Figure 3 (b)). This mode has two orthogonal polarizations and is degenerate. The third mode has a donut shape. However, this donut shaped mode remained the third mode when the radius of the nanowire was varied.

The dispersion relation for the hollow structure was found to be different from the one above. Figure 3 (c) plots the dispersion relation on the hollow nanowire when the ratio between the inner diameter and outer diameter, A, is 0.54. Here, unlike the conventional nanowire waveguide, the effective refractive index becomes low because the inside of the nanowire is air. When the diameter is small, the air hole is smaller than the wavelength, and the profile of the $HE_{11}$ mode is still Gaussian-like. However, as the diameter becomes larger, the electric field changes because the size of the air hole cannot be ignored, and the light cannot localize in the air hole (Figure 3 (d)). Therefore, the effective refractive index of the $HE_{11}$ mode becomes small, and thereby the donut shaped mode becomes fundamental. This kind of mode inversion occurs when the radius is larger than 140 nm. This is a unique feature of this waveguide; such a mode reversal does not occur in common optical fibers. The fundamental mode has the largest effective refractive index, so this means light leakage into the air is small. In other words, in the inversion region, the optical confinement of the donut mode increases. This is an advantageous condition for lasing oscillation. Next, to optimize the inversion conditions, we investigated the effective refractive index difference, $\Delta n$, between the Gaussian-like $HE_{11}$ mode and the donut shaped mode. When this value is negative, the donut shaped mode becomes fundamental. As shown in Figure 4, the minimum value occurs when A=0.725 and r =200 nm. These values are technically feasible, as shown in the SEM in the inset. We also examined the cavity modes. As shown in Figure 5, a cavity mode with an azimuthal polarization was obtained. It was the lowest mode and had a Q factor of over 2000, high enough to achieve lasing oscillation.

## Measurement

The measurement was carried out in a typical micro-photoluminescent (μ-PL) setup at room temperature. The pump laser was 266 nm with a 1-KHz reputation rate and 350-ps pulse width. The standing nanowires were excited from the top. The pump beam was focused by a 0.55 NA objective lens. The emission spectra were measured by 30-cm spectrometer with a liquid nitrogen cooled Si detector. To estimate the emission lifetime, a time-resolved measurement was carried out with a 266-nm femtosecond laser with an 80-MHz reputation rate, a single photon detector, and a time-correlated single photon counting module (see supplemental 2 for details).

## Results

First, the emission lifetimes of the nanowires and hollow nanowires were measured in order to investigate the effect of fabrication damage. Figure 6 shows the emission lifetimes of nanowires with and without hollow structures. The nanowires and hollow nanowires had emission lifetimes of 110 ps and 59 ps, respectively (measurement limit: 32ps). Compared with the previously reported GaN self-catalytic nanowire grown by molecular beam epitaxy[42], the nanowires fabricated by the sublimation method seem to have slightly larger amounts of non-radiative recombination. As can be seen from the SEM image in Figure 2, the surface of the nanowires was somewhat rough, so non-radiative recombination, such as due to tangling bonds, may have been larger than usual. Furthermore, the non-radiative recombination increased in the hollow nanowires due to the increase in surface area. However, as we previously reported, the nanowires had sufficient quality to cause laser oscillation[39,43], and they emitted light well even in a hollow structure as we explain later.

The emission properties of the hollow nanowires were examined by making μ-photoluminescence (μ-PL) measurements at room temperature. Figure 7(a) shows the emission spectrum under the lasing condition. When the excitation intensity was sufficiently strong, a narrow emission line appeared. In order to investigate the characteristics of the laser oscillation in detail, we measured the emission from the hollow nanowires while changing the pump power. Figure 7(b) and (c) show the light-in vs light-out (L-L) characteristics, laser linewidth, and cavity wavelength. A pulse laser with a slow repetition rate was utilized for pumping, so there was no thermal effect or red shift[44]. When the excitation intensity exceeded 10 mJ cm$^{-2}$/pulse, the emission intensity clearly increased, the linewidth narrowed, and a blue shift due to the carrier plasma effect occurred. These results are typical of lasing. Moreover, the lasing threshold of the hollow nanowires was slightly higher than that of nanowires of the same diameter (supplemental 3). This suggests that the threshold is raised by increasing the non-radiative recombination rate.

Figure 8(a) shows the polarization characteristics of this laser mode. From the polar plot in the figure, it can be seen that the emission properties of the nanowire do not exhibit extreme polarization dependence. In general, a nanowire laser oscillation has one polarization component that strongly appears, because the fundamental ($HE_{11}$) mode is uniformly polarized. However as discussed above, the fundamental mode of our hollow nanowire is expected to be axisymmetric polarized. Therefore, the weak polarization dependence indicates vector beam generation. As shown in Figure 7(a), another very small mode exists several nm away on the short wavelength side. This mode spacing between the lasing peak and the small peak corresponds with that of the FEM calculation described in supplemental 4, confirming that the lased mode was donut shaped. In addition, a far field pattern (FFP) was observed (Fig. 8(b)) and it showed a clear donut shaped emission. When the polarizer is rotated, polarized light along the angle of the polarizer can be extracted. Such a pattern is a characteristic of a typical vector beam[33,34]. The FFP was further confirmed in FEM simulations (supplemental 5).

## Summary and outlook

We fabricated a hollow nanowire laser. No other reports have demonstrated such thin nanowires with long hollow structures. This means that our sublimation method is superior to other methods, such as dry etching, for fabricating such high-contrast structures. Since the hollow nanowires stand upright on the sapphire substrate, which is a low-refractive-index material, we were able to achieve laser oscillation by forming a Fabry-Perot cavity. The confirmation of laser oscillation in hollow nanowires is also a first achievement. A detailed modal analysis revealed that our structure has an axisymmetric fundamental mode, and we experimentally confirmed that the structure emitted a vector beam at room temperature. In addition, such laser structures are suitable for integration into reported fibers[37], as they can vertically emit light. This means that they can be applied to optical communications. It may also be useful for trapping nanoparticles near nanowires. In this report, we showed that this structure has an axially symmetrical polarization mode, but it does not have an OAM mode (supplemental 6). However, if we can introduce a spiral structure, it may be possible to have an OAM mode[37,38]. This structure is very interesting as a new direction of research.

## Author Contributions

M.T. measured the optical properties. S.S. and T.T. fabricated the hollow core pattern. M.T., P.H. and T.Y. performed the FEM simulation. B.D., S.V., S.C., and N.Y. carried out the GaN fabrication process. M.T. analyzed the data and wrote the manuscript with contributions from all authors. M.N., A.S, and H.S supervised the entire project. All authors have given approval to the final version of the manuscript.

## Acknowledgements

The authors thank Shinichi Fujiura for helping with the device measurements. They also thank Yoshio Ohki (NTT Advanced Technology Corporation) and Osamu Moriwaki

(NTT Advanced Technology Corporation) for technical support with the FIB fabrication, SEM, and electron-beam lithography.

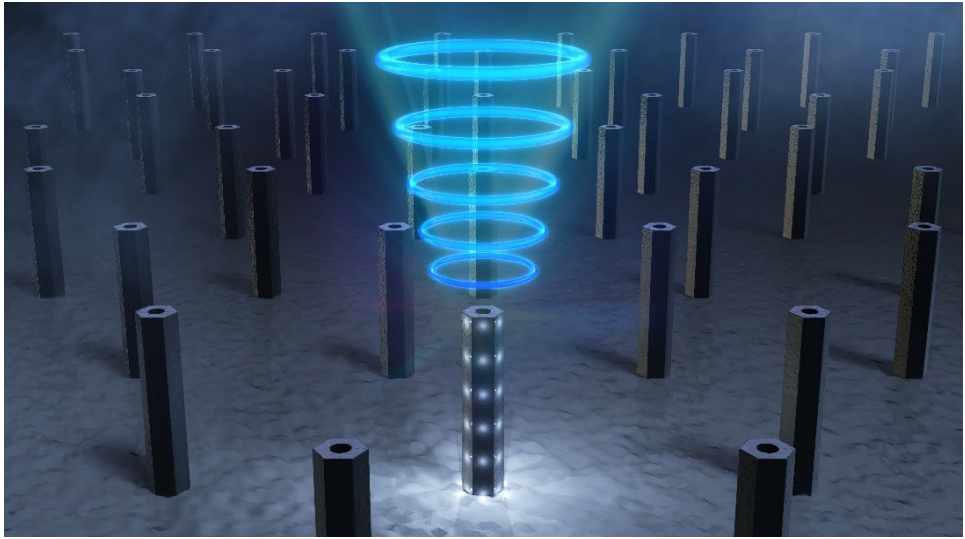

Figure 1 Schematic image of GaN hollow nanowires on sapphire substrate.

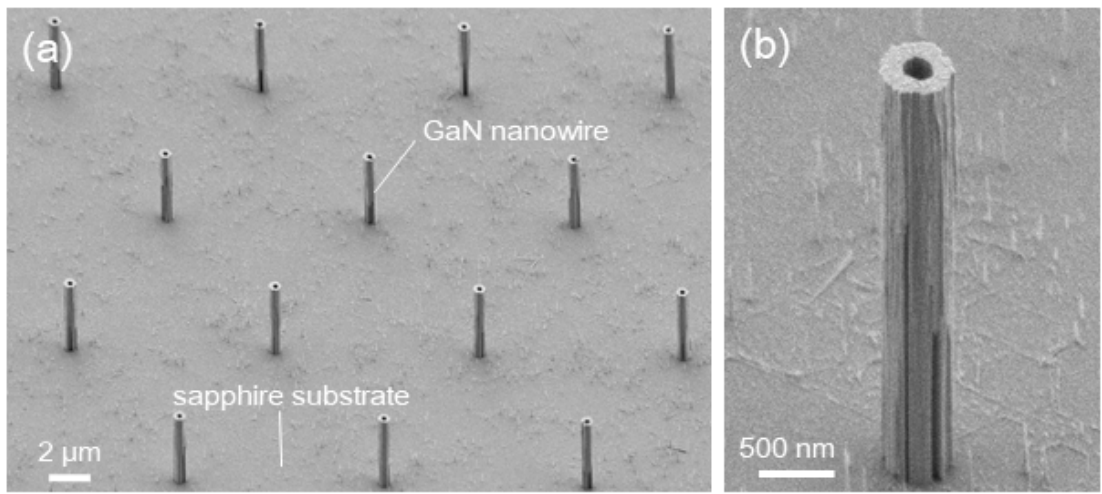

Figure 2 (a) SEM image of GaN hollow nanowires on sapphire substrate. (b) SEM image of a single GaN hollow nanowire.

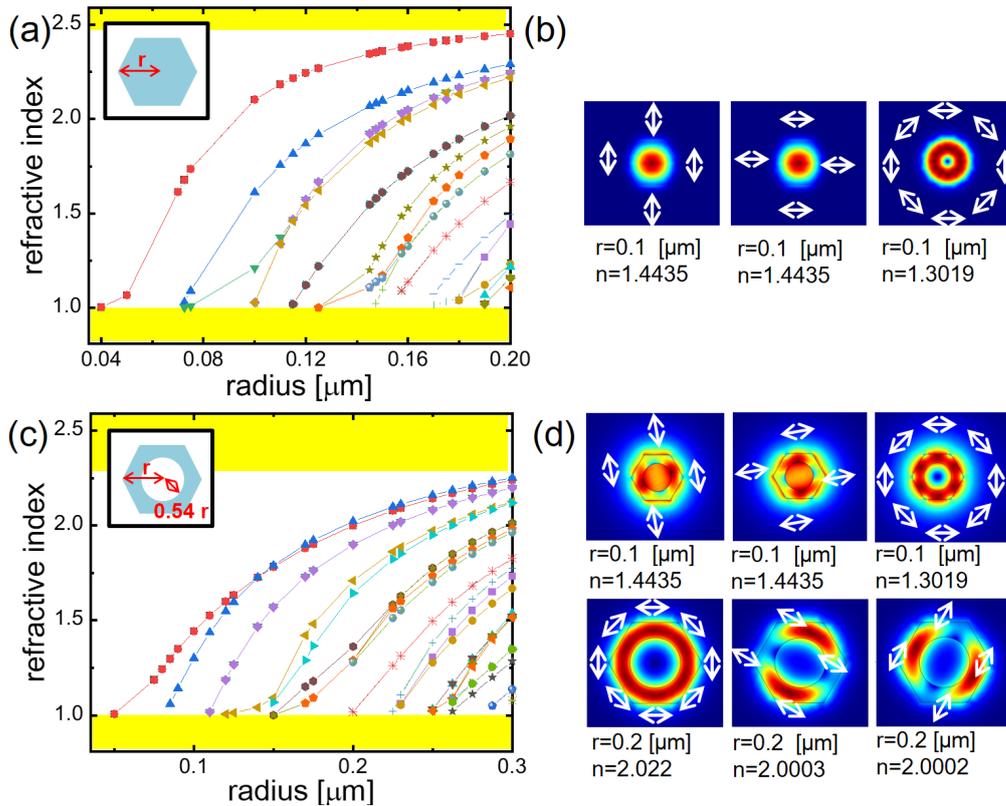

Figure 3 (a) Effective mode refractive indices for an infinitely long GaN nanowire in air versus nanowire diameter. (b) Intensity profiles of the first, second and third modes at 100-nm radius. Arrows show the direction of the electric field. (c) Effective mode refractive indices for an infinitely long GaN hollow nanowire in air versus nanowire diameter. (d) Intensity profiles of the first, second and third modes at 100-nm and 200-nm radius.

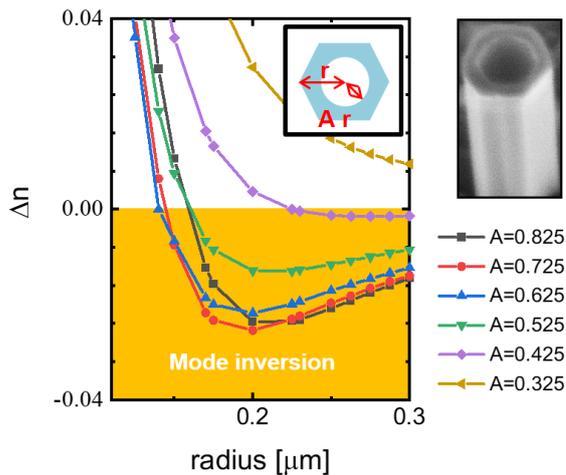

Figure 4 Difference in effective mode refractive index between $HE_{11}$-like modes and donut shaped mode.

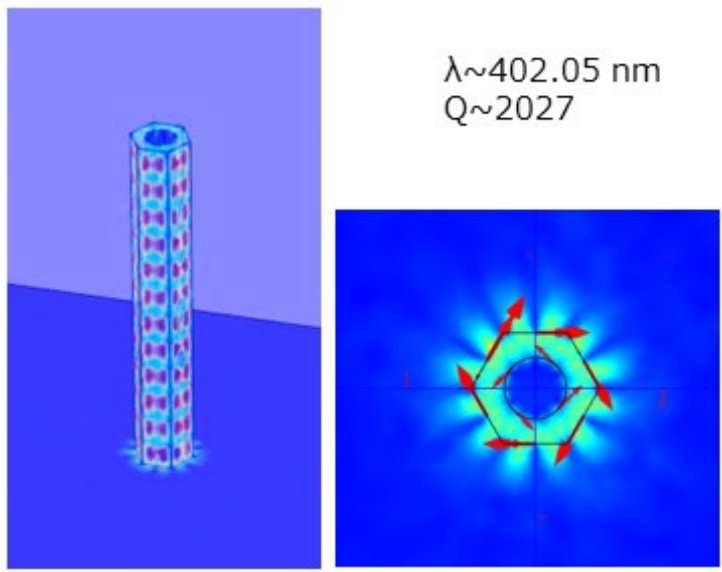

λ~402.05 nm
Q~2027

Figure 5 Cavity mode of a hollow nanowire on sapphire substrate. Red arrows show the direction of the electric field at a moment.

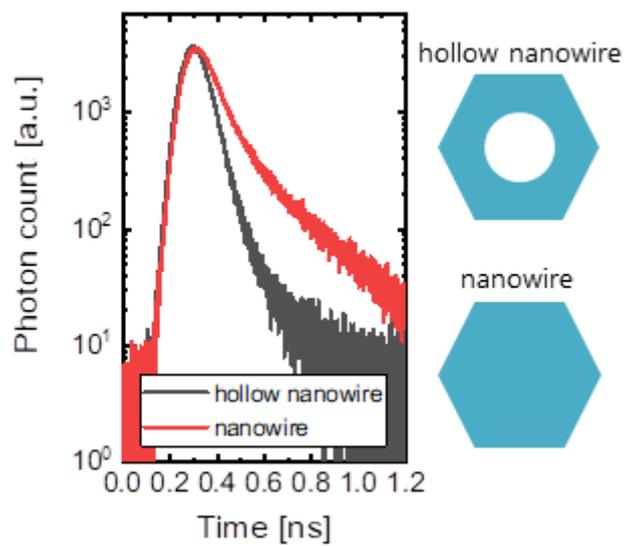

Figure 6 Emission lifetime of nanowire with and without hollow structures

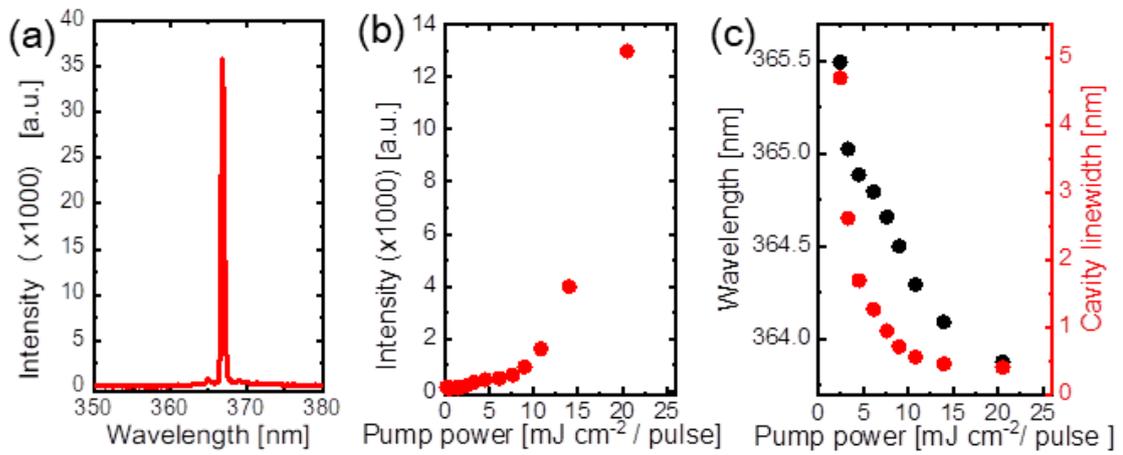

Figure 7 (a) Emission spectrum of a GaN hollow nanowire under strong pumping condition. (b) Light-in versus light-out curve. (c) Wavelength and emission linewidth of the spectrum.

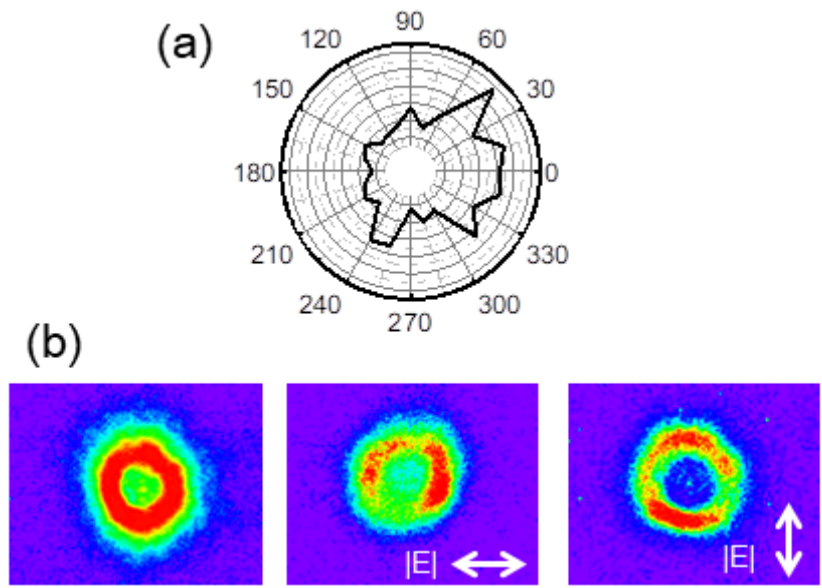

Figure 8 (a) Polar plot of the emission (b) Far field pattern with and without polarizer. Arrows show the direction of the electric field.

# (Supplementary information)
# Vector beam generation from standing hollow GaN nanowire lasers on sapphire substrate

Masato Takiguchi[1,2*], Sylvain Sergent[1,2], Benjamin Damilano[3], Stéphane Vézian[3], Sébastien Chenot[3], Nicole Yazigi[3], Taiki Yoda[1,2], Peter Heidt[2], Tai Tsuchizawa[1,4], Hisashi Sumikura[1,2], Akihiko Shinya[1,2], and Masaya Notomi[1,2]

[1]Nanophotonics Center, NTT Corp., 3-1, Morinosato Wakamiya, Atsugi, Kanagawa 243-0198, Japan
[2]NTT Basic Research Laboratories, NTT Corp., 3-1, Morinosato Wakamiya, Atsugi, Kanagawa 243-0198, Japan
[3]Université Côte d'Azur, CNRS, CRHEA, Rue B. Grégory, Valbonne 06560, France
[4]NTT Device Technology Laboratories, NTT Corp., 3-1, Morinosato Wakamiya, Atsugi, Kanagawa 243-0198, Japan

E-mail: masato.takiguchi@ntt.com

## S1. Cross section of a hollow core nanowire

To confirm whether a hollow structure had been fabricated, we cut a nanowire by using a Ga ion focused ion beam (FIB) system. The scanning electron microscope (SEM) image on the left shows a 4-um-long nanowire lying on a sapphire substrate. The SEM image on the right shows a cross section of this nanowire cut with the FIB. As can be seen in the figure below, the hole went all the way through the nanowire.

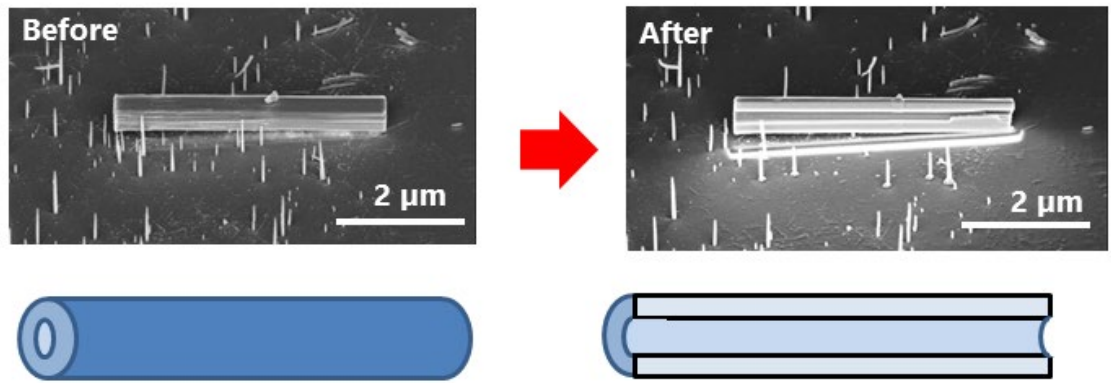

Figure S1. SEM image of hollow nanowire before and after FIB cutting.

## S2. Measurement setup

Photoluminescence (PL) and time-resolved measurements were carried out using a typical setup with a 266-nm pump laser that can optically excite GaN (Figure S2). The sample's emission was measured with a 30-cm spectrometer equipped with a liquid-nitrogen-cooled Si CCD camera and a single photon detector. The measurement system detected the light emitted from the top of the nanowires. To separate the pump laser (226 nm) and the emission (about 360 nm), we used a dichroic mirror and a longpass filter. For the time resolved measurement, we initially tried a femtosecond laser with a repetition frequency of 80 MHz. Our nanowires did not achieve lasing oscillation because of thermal effects caused by the high pulse repetition frequency of the pump laser. Therefore, we excited our nanowire with a lower pulse repetition frequency (1 KHz). The emission was collected by 0.55 NA objective lens which had a large NA in the UV region.

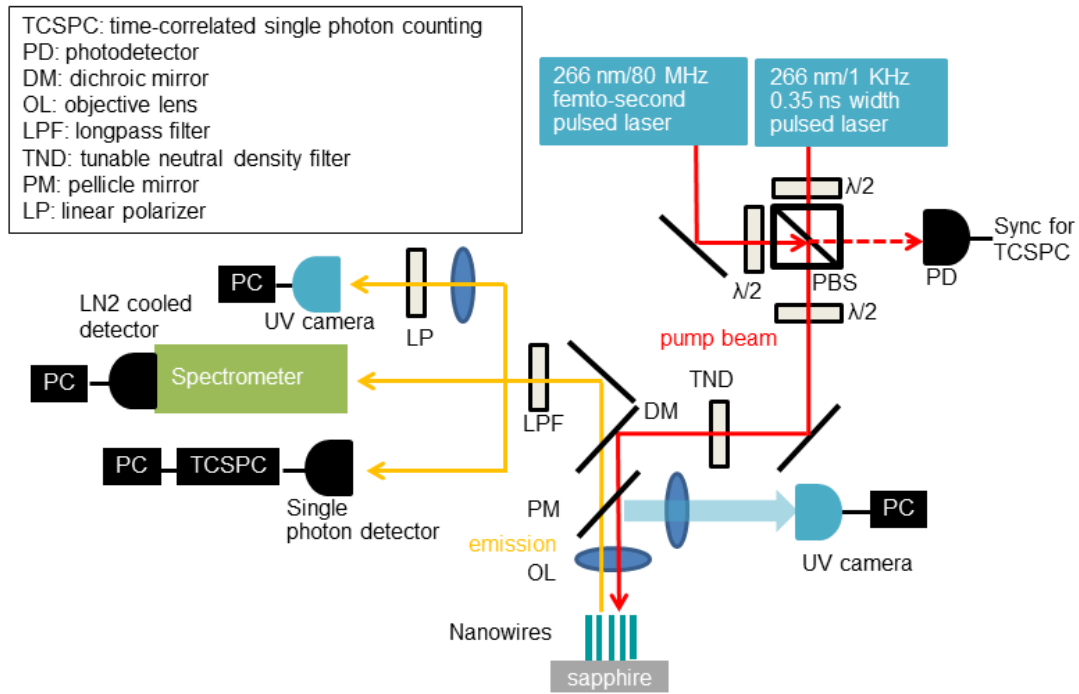

Figure S2. Schematic measurement setup

## S3. Photoluminescence measurement of nanowires

Figures 3S(a) and (b) show the light-in versus light-out (L–L) curve and the emission spectrum after lasing under the strong pumping condition. The emission spectrum is sharp at a pump power around 7 mJ cm$^{-2}$ / pulse. This threshold power is smaller than that of the hollow nanowire laser. This indicates the nonradiative recombination rate is lower than in the hollow nanowire. However, this nanowire can easily show multimode lasing because many modes exist in this condition. For this reason, the emission intensity in Figure 3(a) fluctuated due to mode hopping.

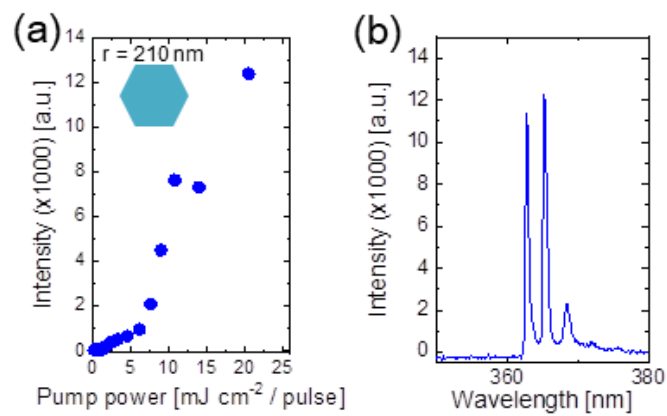

Figure S3. (a) Light-in versus light-out curve of a nanowire without an air hole. (b) Emission spectrum under strong pumping condition.

## S4. Cavity mode analysis

Figure S4 (a) shows a 3D FEM simulation of the cavity modes in a hollow nanowire. These cavity modes include the fundamental one that generates the vector beam and higher order ones. Figure S4 (b) shows the experimental results, which consist of the same data as in Fig 7(a). As shown, there are two peaks. One is the lasing mode and the other is a non-lasing mode (the part pointed to by the arrow). The mode spacing (a few nm) is consistent with the spectral spacing obtained from the experimental results. Additionally, the fundamental mode has a higher Q and is more advantageous for lasing oscillation. However, if the nanowire modes are unstable under high-power excitation, higher-order modes could show lasing (Fig S4 (c)). In this case, the higher order mode has a single polarization characteristic.

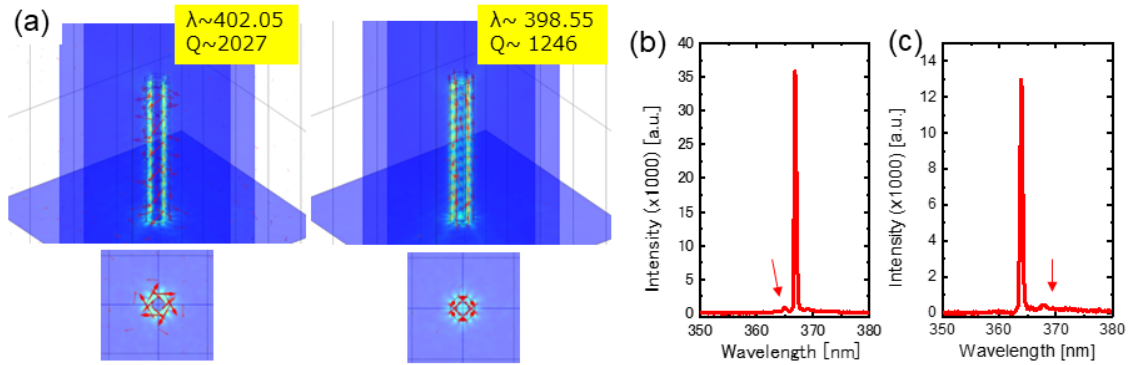

Figure S4. (a) Cavity modes of a hollow nanowire. (b) Lasing spectrum (fundamental mode). (c) Lasing spectrum (high order mode)

## S5. Far field pattern

To estimate the far field pattern (FFP), we carried out a 3D finite element method (FEM) simulation. First, we checked the waveguide mode of the hollow nanowire. The upper part of Figure S5 shows the absolute value of the electric field near the nanowire surface for different eigenvalues. The bottom part of Figure S5 shows images of the FFPs obtained by Fourier transform. A large NA objective lens (NA 0.71) is needed to collect the complete electric field profile for the donut shaped pattern. The 0.55 objective lens of the experiment does not collect all the emission, which may reduce the contrast of the signal (In the experiment, we change the focus to collect all the light). As shown, other modes based on $HE_{11}$ have larger spread angles from nanowires. This spread may help to distinguish between the donut-shaped ($TE_{01}$) and $HE_{11}$ modes.

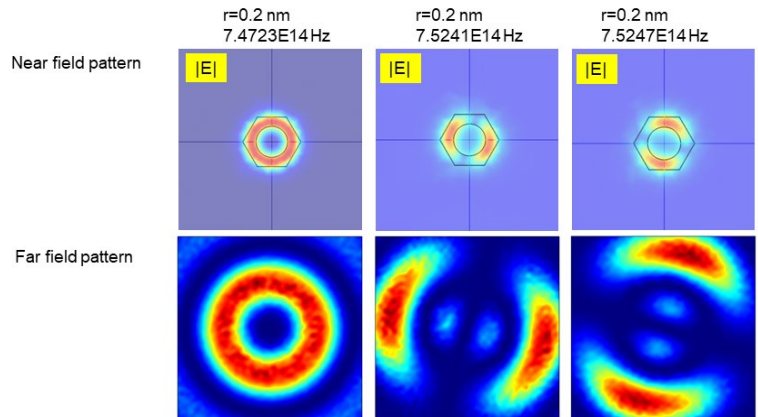

Figure S5. Near field patterns and far field patterns of a hollow nanowire

## S6. Orbital Angular Momentum

To determine whether the fundamental mode has momentum, we examined the polarization components when changing the phase. As shown in Figure S6, the directions change from clockwise to counterclockwise. This means that it does not have a polarization in a specific direction when averaged over time. Therefore, this mode does not have OAM. The direct calculation shows that when the light escaping the nanowire satisfies the paraxial approximation, the transverse components of the OAM can be neglected, and only the longitudinal component, $J_z$, is substantial. The normalized OAM, $J_z$, is given by the following formula [M V Berry et a., 2005, J. Opt. A: Pure Appl. Opt. **7** 685]:

$$\frac{J_z}{W} = \frac{Im \int_0^\infty \int_0^{2\pi} E^*(r,\phi,z)\frac{\delta}{\delta\phi}E(r,\phi,z)r dr d\phi}{\int_0^\infty \int_0^{2\pi} |E(r,\phi,z)|^2 r dr d\phi},$$

where W is optical power, E is electric field, and r, φ, z are cylindrical coordinates. The OAM was calculated at a transverse plane 50 nm above the nanowire facet and zero OAM was confirmed for the near-field. By conservation of OAM in free-space, the far-field OAM is likewise zero.

**phase**

| **0** | **$\pi/2$** | **$\pi$** | **$3\pi/2$** |
|---|---|---|---|

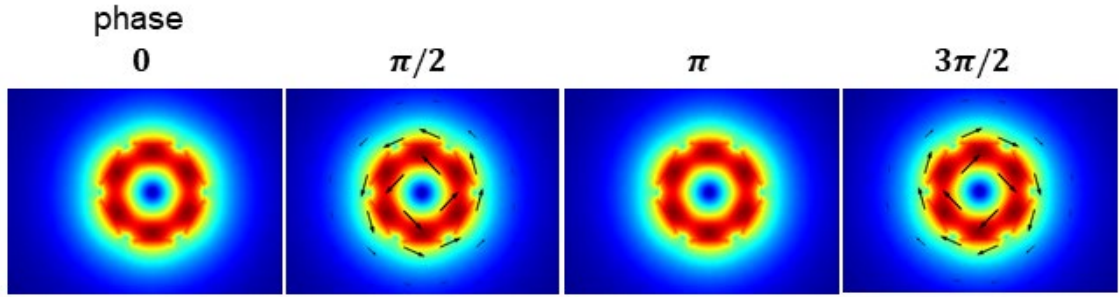

Figure S6. Polarization direction of the electric field of a hollow nanowire.